# Direct Inversion of Digital 3D Fraunhofer Holography Maps

Sergey G. Podorov*[1,2]
and Eckhart Förster,[2]

[1]x-ray-soft.ru,Emwa,169200,Russia
[2]Helmholtz Institute Jena, Fröbelstieg 3, 07743, Germany
*xtris@mail.ru

**Abstract:** The Differential Fourier Holography (DFH) gives an exact mathematical solution of the inverse problem of diffraction in the Fraunhofer regime. After the first publication [1] the Differential Fourier Holography was successfully applied in many experiments to obtain amplitude and phase information about two-dimensional (2D) images. In this article we demonstrate numerically the possibility to apply the DFH also for investigation of unknown 3D Objects. The first simulation is made for a double-spiral structure plus a line as a reference object.

## 1. Introduction

Coherent Diffractive Imaging (CDI) is one of interesting research methods in optics that has dynamically developed in recent years. In the CDI experiments an unknown object is illuminated with a coherent plane wave and far-field diffracted intensity is registered by a detector. The CDI problem is how to reconstruct the optical properties of the unknown object by using the diffracted intensity data. As phase information of diffracted waves is not registered by the detector, such reconstruction is a difficult mathematical problem in the general case. As the complex conjugated wave gives identical intensity on the detector, the corresponding inverse problem has at least two different solutions. Mathematically it means that the inverse problem is ill posed and corresponding numerical solution algorithms have generally a bad or slow convergence. The problem may have even no solutions in presence of noise or in case of a small intensity level on the detector. Nevertheless the CDI problem has various exact solutions in several cases. After discovery of the Gabor Holography [2], the first variants of the Fourier Holography (FH) [3] have been developed. In the Fourier Holography the phase problem was solved by introducing of a small reference hole near the investigated object. In this case we can obtain amplitude-phase information about the object by direct application of the Fourier Transform (FT) to the measured intensity from the CCD camera. The reconstruction is in this case a very fast numerical procedure. Iterative algorithms do not need reference holes and later they have been considered as standard routines to solve inverse problems of diffraction [4, 8]. The CDI problem is very close to the diagnostic of heterostructures by X-ray diffraction. The problem is how to determine crystal lattice parameter distribution in multilayered crystal films from corresponding X-ray rocking curves. For plane structures the problem is mathematically identical to the CDI problem, but it is one-dimensional. In this case [5], only average characteristics were determined [6]. There are also different direct methods of visualization of weakly absorbing objects, which use different special optical elements. For example, a system of asymmetrically cut crystals [7, 9] is applied as so called monochromator and analyzer. The object to study is placed between the crystals, for more details see [25]. It is evident, that the spatial resolution of the FH method is limited not only by the diffraction theory, but also by the geometrical conditions as the size of the used pinhole. By differentiation of various objects it is possible to create mathematically Dirac delta-like objects as by using a physical pinhole. Firstly such techniques were applied to determine crystal lattice strain in heterostructures with one-dimensional variation of their chemical composition with crystal depth [10, 11]. Later the Differential Fourier Holography method was used in the 2D case [1, 13]. In experiments the use of DFH technique means simply that object to study must be placed near to specially formed slits or the reference object with sharp corners. In an experimental work Zhu et al**.** [14] provide a comparison of different lens less imaging methods and showed that the DFH method gave higher spatial resolution as the other CDI methods. There are also different attempts [12] to achieve nano-resolution by special X-ray crystal optics with two bent crystals. Enders et al. [15] have shown that each amplitude-phase object, that has a large difference in its optical properties relative to the substrate, can be used as an extended reference object for DFH. It was shown in [16] that the DFH may be used in two different geometries. The rectangular reference object may be a mask or a hole with an object to study near to a corner of the reference object. Sumriddetchkajorn et al. [17] used the DFH method for the quality analysis of sub-millimeter thick bars. To solve the inverse problem, other authors [18, 19] developed different direct methods. Gauthier et al. [20] and Nishino et al. [21] have shown possibilities to reconstruct objects by the DFH method using only a single shot of a femtosecond pulse. Lan et al. [23] demonstrated that the HERALDO method in a general formulation by Guizar-Sicairos et al. [13] does not give the exact solution of the inverse problem in general cases, but such solution may be used as a first step for the further iterative procedure. Capotondi et al. [24] used a combination of DFH and methods of lens less imaging for X-ray microscopy.

## 2. Method of Measurements and Theory of 3D Holography

After successful applications of CDI methods in 2D case many researchers are trying to extend 2D CDI techniques for Three-Dimensional objects. Thus Gulden et al. [22] have tried to reconstruct "three-dimensional structure of a single colloidal crystal grain", but to reconstruct the full structure of the grain was impossible. The 3D diffracted intensity is obtained from a set of 2D images from CCD detector by rotating the sample and finally by combination and interpolation of such images into the full 3D intensity (for more details see [22]). To solve the 3D CDI problem different iterative methods are used identical to the 2D CDI case. By iterative methods, researchers have to make many attempts with different initial values and compare many numerical results after 2000-5000 iterations of each. Due to bad and instable convergence of iterative procedures as well as their dependence on start approximation, to find best result is a very difficult problem what needs long calculation times. It is obvious that corresponding 3D CDI problem may take much more time for solution by iterative methods. Due to this reason it is very attractive to apply the fast methods of Differential Fourier Holography for 3D cases.

Usually we radiate a small sample with plane waves without sample rotation and take a diffracted image of the sample on a plane. The image is a 2D Fourier transform of the studied sample (see Fig. 1). If we will rotate the sample over the axe with small angle step and collect all corresponding 2D images in one 3D set, then we obtain a square of modulus of 3D Fourier Transform of the sample (see Fig. 2). The set of the data will be presented initially in cylindrical coordinate system and must be interpolated in a 3D Cartesian coordinate system more suitable for numerical computations. So we obtain 3D map of diffracted intensity and our problem is to reconstruct 3D structure of the sample analyzing this 3D intensity map.

Generally, it is possible to apply existing iterative methods of 2D CDI for solution of the 3D inverse problem, but we expect sufficient problems with numerical computations in this way. To apply already existing method of differential Fourier Holography is more attractive. If you place suitable 3D extended object near the sample of study, the reconstruction process may be made already in one step by taking 3D Fourier transform of corresponding 3D map.

We describe shortly mathematics of our method. Here, as in 1D or 2D case we can write also in 3D case for the diffracted intensity I:

$$I(\mathbf{s}) \sim | FT(\chi(\mathbf{r})) |^2 , \qquad (1)$$

where FT – 3D Fourier Transform operator, $\chi(\mathbf{r})$ – complex susceptibility of the illuminated objects and **s** – 3D scattering vector. The inverse problem of diffraction is how to determine the complex susceptibility from known measured intensities that satisfy Eq. (1).

We show that for the 3D imaging case, the method of DFH [1] or the combination of DFH and iterative methods as shown for the 2D case in [24] for 2D case may be used for successful reconstructions of sample objects.

The solution method of such problems is as follows – in addition to the unknown object we place a reference object $\chi_{\text{reference}}(\mathbf{r})$ with certain properties:

$$\mathbf{A}\, \chi_{\text{reference}}(\mathbf{r}) = \sum \pm \delta(\mathbf{r} - \mathbf{a}_i), \tag{2}$$

here $\mathbf{A}$ is the differential operator of first, second or third order. If the reference object is a line in direction of vector $\mathbf{p}$, then the operator $\mathbf{A}$ makes the differentiation in the direction of $\mathbf{p}$ – the vector parallel to the line:

$$\mathbf{A}\, \chi_{\text{reference}}(\mathbf{r}) = (\mathbf{grad}\, \chi_{\text{reference}}(\mathbf{r}), \mathbf{p}) \tag{3}$$

To illustrate our DFH method for 3D imaging, we take a 3D structure like a parallel double-spiral as test object, placed near the line object as the reference. Then we perform the FT of the structure and register their "scattered intensity". The reconstruction of 3D images in this way is as follows:

$$\frac{\partial}{\partial \vec{\mathbf{p}}} \iiint_\Omega I(\vec{\mathbf{s}})\exp(i\vec{\mathbf{s}}\vec{\mathbf{r}})ds = (\vec{\mathbf{p}}, \mathbf{grad})\iiint_\Omega I(\vec{\mathbf{s}})\exp(i\vec{\mathbf{s}}\vec{\mathbf{r}})ds =$$
$$= i\iiint_\Omega (\mathbf{p},\mathbf{s})I(\vec{\mathbf{s}})\exp(i\vec{\mathbf{s}}\vec{\mathbf{r}})ds \tag{4}$$

Otherwise

$$\iiint_\Omega I(\vec{s})\exp(i\vec{s}\vec{r})ds \propto \iiint_V \left(\chi_{ref}(\vec{r}') + \chi_{obj}(\vec{r}')\right)\left(\chi_{ref}(\vec{r}'+\vec{r}) + \chi_{obj}(\vec{r}'+\vec{r})\right)d\vec{r}' \tag{5}$$

Combine (4-5) together we obtain:

$$\iiint_\Omega (\mathbf{p},\mathbf{s})I(\vec{\mathbf{s}})\exp(i\vec{\mathbf{s}}\vec{\mathbf{r}})ds \propto$$
$$\propto \iiint_V \left(\chi_{ref}(\vec{\mathbf{r}}') + \chi_{obj}(\vec{\mathbf{r}}')\right)\frac{\partial}{\partial \vec{\mathbf{p}}}\left(\chi_{ref}^*(\vec{\mathbf{r}}'+\vec{\mathbf{r}}) + \chi_{obj}^*(\vec{\mathbf{r}}'+\vec{\mathbf{r}})\right)dx'dy'dz' \propto$$
$$\propto \iiint_V \left(\chi_{ref}^*\delta(\vec{\mathbf{r}}'+\vec{\mathbf{r}}-\vec{\mathbf{a}}) - \chi_{ref}^*\delta(\vec{\mathbf{r}}'+\vec{\mathbf{r}}-\vec{\mathbf{b}}) + \frac{\partial}{\partial \vec{\mathbf{p}}}\chi_{obj}^*(\vec{\mathbf{r}}'+\vec{\mathbf{r}})\right)\times \tag{6}$$
$$\times\left(\chi_{ref}(\vec{\mathbf{r}}') + \chi_{obj}(\vec{\mathbf{r}}')\right)dx'dy'dz' \propto$$
$$\propto \left(\chi_{ref}(\vec{\mathbf{a}}-\vec{\mathbf{r}}) + \chi_{obj}(\vec{\mathbf{a}}-\vec{\mathbf{r}})\right)\chi_{ref}^* - \left(\chi_{ref}(\vec{\mathbf{b}}-\vec{\mathbf{r}}) + \chi_{obj}(\vec{\mathbf{b}}-\vec{\mathbf{r}})\right)\chi_{ref}^* +$$
$$+ \textit{mirrored c.c.} + \textit{noise}$$

Differentiation of the extended reference object in direct space gives delta functions at corners of the reference object. The convolution of obtained delta functions with illuminated objects gives a set of exact solutions of the inverse problem with additional noise. Thus taking the integral (4) we reconstruct 3D image of the sample in one computational step.

We proof numerically our solution method. We took a short line as a reference object for the simulation and a 3D parallel double dashed spiral as test object. In Fig. 3 a) we show the reference object as a line in the direction OX and placed it near to the double-spiral object. By using Eq. 1 we simulate the scattered 3D intensity (Fig. 1 c). In Fig. 3 b) we can see one-step reconstruction of the sample object (4 images) that we obtained from Fig. 3 c) by application of the operator **A** to the inverse Fourier Transform of the 3D intensity according Eq. (4).

As the result in Fig. 3 b) we see four images of the sample object together with the reference line and some noise. The noise results from the autocorrelation of double-spiral object with the reference line. For visualization of Fig. 3 a) – 1 c) authors have used routines SHADE_VOLUME() and TV, POLYSHADE() of IDL 8.2.

The method has the following numerical properties. All calculations must be made with double precision, also the 3D version of FFT (calculations lose numerical accuracy and do not give the needed results in the case of numeric data with single precision). The conventional picture formats such as GIF and TIFF are not suitable to save the experimental intensity data, as they have a small dynamic range what is not sufficient for good reconstruction.

Many researchers have been very skeptical about of application possibility of Differential Fourier Holography. They have the opinion that measured experimental intensity is sufficiently dropping especially in case of small intensity and in such case differentiation of the Fourier Transform of the measured intensity is not mathematically correct. We remark that all measured maps of intensity are from $L_2$ space and Fourier Transforms of functions from $L_2$ space are differentiable. Mathematically we do not need to differentiate anything, according to right part of Eq. (4), see Eq. (7), we have to take numerically only the Fourier Transform of the modified map - without any numerical differentiation.

$$object + c.c. mirrored\ object + noise \propto \iiint_\Omega (\mathbf{p},\vec{\mathbf{s}}) I(\vec{\mathbf{s}}) \exp(i\vec{\mathbf{s}}\vec{\mathbf{r}}) ds \qquad (7)$$

The spatial resolution of our method depends on several factors. The distance to the detector and the numerical aperture set a theoretical limit for the spatial resolution of the scheme. Another factor is the angular step of collection of the 2D images into the 3D map. Perfectness and geometry of the reference object is the next factor. The last factor is the accuracy of the diffracted intensity measurements.

The dropping of the intensity at low level has surely a negative effect on the reconstruction as it was discussed by Podorov et al. [1] already in the first paper related to DFT and later by other authors. Other reconstruction methods are not working too in the case of low intensity data with large noise. Such situation is expected as the inverse problem is mathematically ill posed problem due to knowing theorem as there are many possible solution of the problem, for example direct and complex conjugated and mirrored images.

Generally, different 1D, 2D and 3D objects with sharp corners may be used as reference object. What by application of a suitable differential operator produces 3D Dirac delta-functions and allows us to produce reconstructions of the 3D image of studied samples.

With above described conditions, the new method of 3D CDI inversion is very fast and gives several exact solutions of objects to study, i.e. direct and complex conjugated mirrored objects.

## 3. Conclusion

In this article we successfully demonstrated the applicability of Differential Fourier Holography method [1, 13-16, 20-21] for the case of 3D imaging [22]. For demonstration as the reference object a line was taken. We underline that as the reference many other 3D objects may be taken that by differentiation give a set of Dirac-delta functions at their corners positions. Our method can be applied too as a start approach for iterative procedures as in [24] to reduce the reconstruction time.

Generally, our method is fast, numerical one-step reconstruction procedure and may be applied not only with coherent X-rays radiation, but with any other coherent electromagnetic radiation under conditions of the Fraunhofer diffraction.

**4. Acknowledgments**

Sergey G. Podorov acknowledges the Helmholtz Institute Jena and the GSI Darmstadt for financial support of this work.

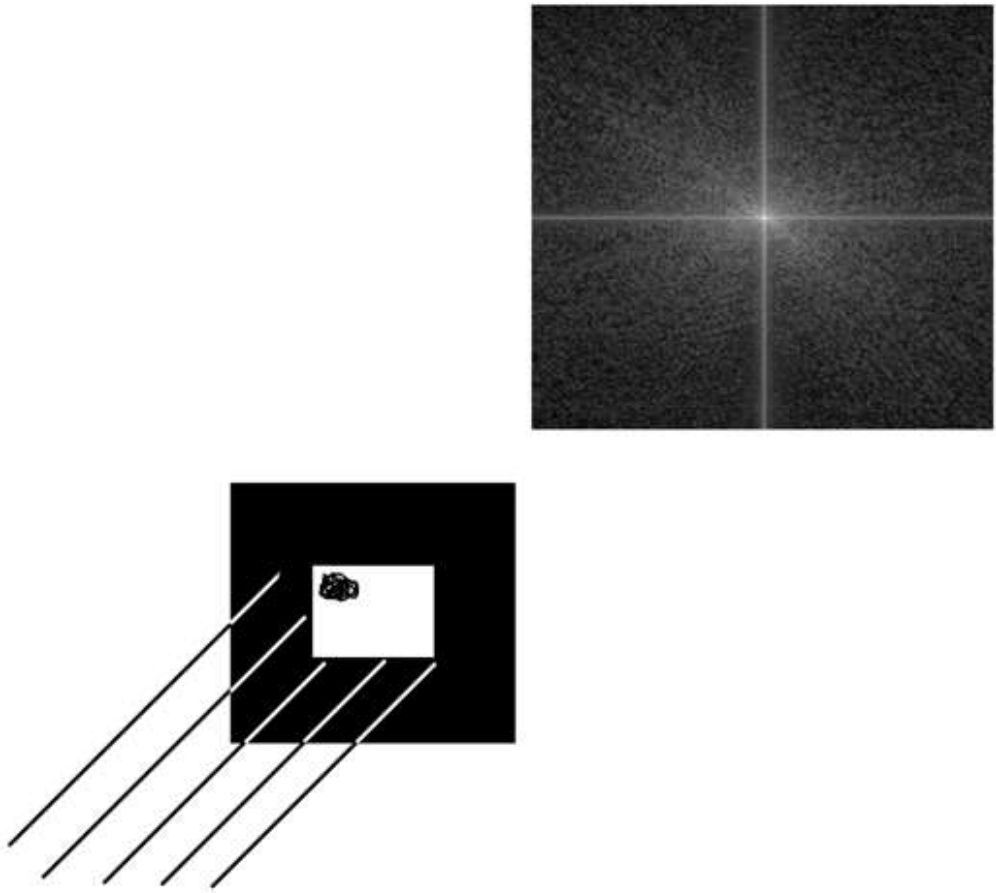

Fig. 1 Illuminated object within a mask irradiated by plane waves and its Fraunhofer 2D diffraction map.

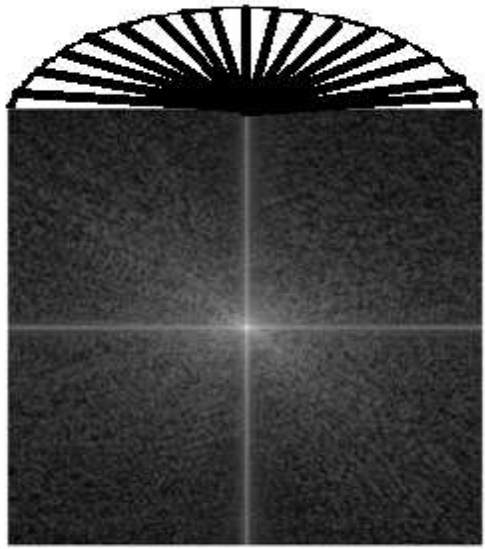

Fig. 2 Building of 3D Fraunhofer diffraction map by collection of a set of 2D diffraction maps in a one 3D model with corresponding interpolation between 2D images.

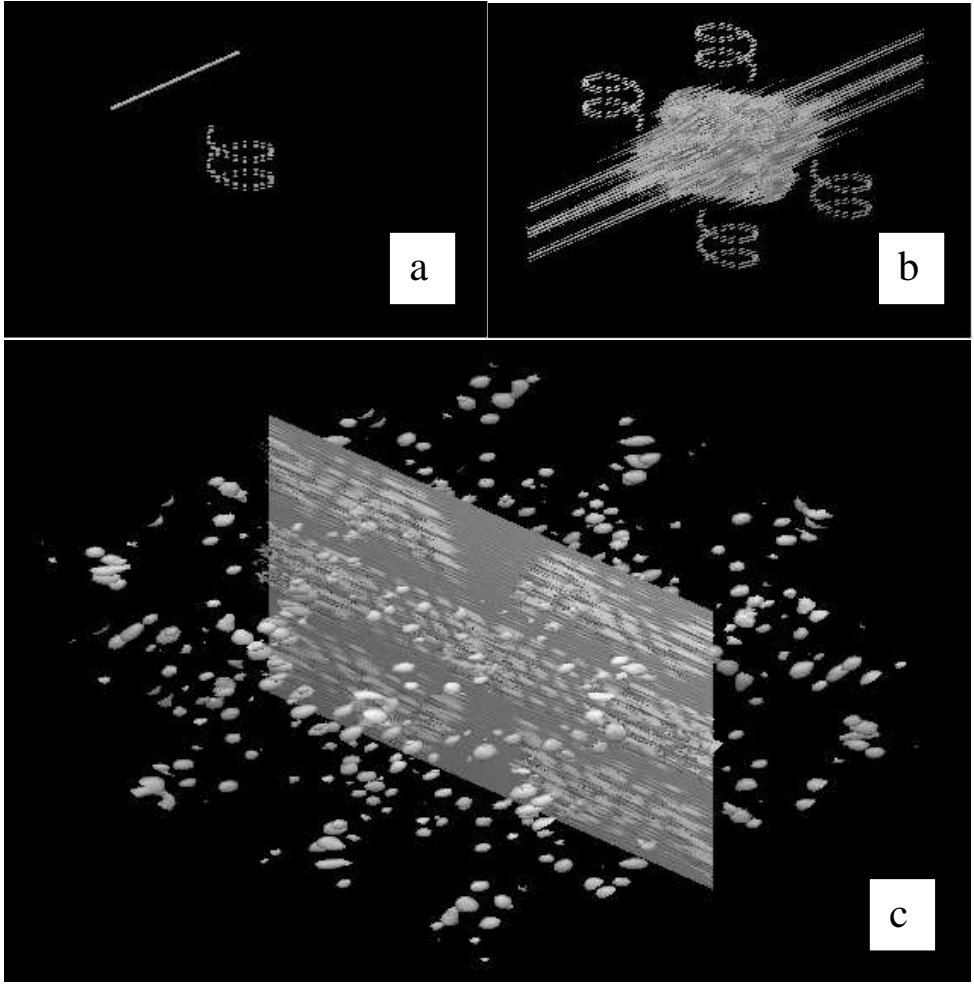

Fig. 3 a – Parallel double-spiral dashed object and a line as reference object, b - Reconstructed by Differential Fourier Holography double spiral objects and their mirrored images.
 c - Visualization of scattered 3D intensity map.